\documentclass[aps,twocolumn,showpacs,preprintnumbers,floats]{revtex4}

\usepackage{graphicx}
\usepackage{dcolumn}
\usepackage{bm}
\usepackage{epsfig}

\begin{document}

\title{ $\eta$-meson
       in nuclear matter}
\author{
 X. H. Zhong,$^1$\footnote{E-mail: zhongxianhui@mail.nankai.edu.cn}
 G. X. Peng,$^{3,2}$\footnote{E-mail: gxpeng@ihep.ac.cn}
 Lei Li,$^{1}$ and
 P. Z. Ning$^{1}$\footnote{E-mail: ningpz@nankai.edu.cn}
       }
\affiliation{
 $^1$Department of Physics, Nankai University, Tianjin 300071, China\\
 $^2$Institute of High Energy Physics,
       Chinese Academy of Sciences, Beijing 100039, China\\
 $^3$China Center of Advanced Science and Technology
       (World Lab.), Beijing 100080, China}

\begin{abstract}
The $\eta$-nucleon ($\eta$N) interactions are deduced from the
heavy baryon chiral perturbation theory up to the
next-to-leading-order terms. Combining the relativistic mean-field
theory for nucleon system, we have studied the in-medium
properties of $\eta$-meson. We find that all the elastic
scattering $\eta$N interactions come from the
next-to-leading-order terms. The $\eta $N sigma term is found to
be about 280$\pm$130 MeV. The off-shell terms are also important
to the in-medium properties of $\eta$-meson. On application of the
latest determination of the $\eta$N scattering length, the ratio
of $\eta$-meson effective mass to its vacuum value is near
$0.84\pm0.015$, while the optical potential is about $-(83\pm5)$
MeV, at the normal nuclear density.
\end{abstract}

\maketitle

\section{Introduction}
The studies of meson-baryon interactions and the meson properties
in nuclear medium are interesting subjects in nuclear physics. The
pion-nucleon/pion-nucleus and kaon-nucleon/kaon-nucleus
interactions have been much studied, both theoretically and
experimentally. Due to the lack of eta beams, the
$\eta$-nucleon/$\eta$-nucleus interaction
is still not as clear as that of
the pion-nucleon/pion-nucleus and kaon-nucleon/kaon-nucleus.
Since the $\eta$-nucleus quasi-bound
states were first predicted by Haider and Liu \cite{b1} and Li
\emph{et al.} \cite{b2}, when it was realized that the
$\eta$-nucleon interaction is attractive, the study of the
$\eta$-nucleus bound states has been one of the focuses in nuclear
physics \cite{b4,b5,b6,b7,b8,b9,b10,b11,b12}.

The key point for the study of $\eta$-nucleus bound states is the
$\eta$\ nuclear optical potential. There have been some works in
this field.
Waas and Weise studied the s-wave interactions of $\eta$-meson in
nuclear medium, and got a potential $U_{\eta}\simeq -20$ MeV
\cite{b3}. Chiang {\sl et al}.\ \cite{op} gave $U_{\eta}\simeq
-34$ MeV by assuming that the mass of the $N^*(1535)$ did not
change in the medium. Tsushima {\sl et al}.\ predicted that the
$\eta$-meson potential was typically $-60$ MeV using QMC model
\cite{efm}. Inoue and Oset also obtained $U_{\eta}\simeq -54$ MeV
with their model \cite{op1}. Obviously, there are model
dependencies in describing the in-medium properties of
$\eta$-meson. Therefore, further studies are needed. In this
paper, firstly we deduce the $\eta$N interactions from chiral
perturbation theory; then combining the relativistic mean-field
theory for nucleon system, we will study the properties of eta
meson in uniform  nuclear matter.

The relativistic mean field theory (RMF) is one of the most
popular methods in modern nuclear physics. It has been successful
in describing the properties of ordinary nuclei/nuclear matter
and hyper-nuclei/nuclear matter\cite{rmf,Lalazissis97prc55}.

On the other hand, the chiral perturbation theory (ChPT) was first
applied by Kaplan and Nelson to investigate the in-medium
properties of (anti)kaons \cite{chiral01}. Some years later, an
effective chiral Lagrangian in heavy-fermion formalism
\cite{chiral1} was also introduced to study the
kaon-nuclear/nucleon interactions or kaon condensation
\cite{chiral,CHL,chiralpp}. The advantage of using the
heavy-fermion Lagrangian for chiral perturbation theory was
clearly pointed out in Ref.\ \cite{chiral1}. Compared with the
previous chiral perturbation theory \cite{chiral01}, the
outstanding point  
in Refs.\ \cite{chiral,CHL,chiralpp} is that  additional
next-to-leading-order terms, i.e., off-shell terms, are added to
the Lagrangian.
%
The additional  terms are essential for a correct description of
the
KN interactions. 
%

The chiral perturbation theory also had been used in the study of
$\eta$-meson in-medium properties in Ref.\ \cite{b3,op1}, where
only the leading-order terms were kept in the calculations. Given
that the higher order terms, e.g., off-shell terms, are important
to the $\eta$N interactions, and they have not been included in
the previous studies for the $\eta$N interactions with chiral
perturbation theory, we have, in the present work, studied the
$\eta$N interactions  with the heavy-baryon chiral perturbation
theory up to the next-to-leading-order terms. Combining the RMF
for nuclear matter, we obtain the in-medium properties of
$\eta$-meson. Comparing our results with the previous results
(with only leading-order terms), we find that the
next-to-leading-order terms are important to the calculations
indeed. The $\eta$-nucleon sigma term is found to be 280 $\pm$\
130 MeV. The ratio of $\eta$-meson effective mass to its vacuum
value is $0.84\pm0.015$, while depth of the optical potential is
$-(83\pm 5)$ MeV, at the normal nuclear density. The large
uncertainty in the sigma term $\Sigma_{\eta N}$ does not affect
the results significantly in low density region, varying by about
8 MeV at normal nuclear density.

The paper is organized as follows. In the subsequent section, the
effective chiral Lagrangian density we used is given, the
effective Lagrangian for $\eta$N interactions is derived, and the
coefficients for the sigma and off-shell terms are determined.
Then, combining the RMF for nucleons, we obtain the $\eta$-meson
energy, effective mass, and optical potential in nuclear matter in
Sec.\ \ref{mp}. We present our results and discussion of  the
$\eta$-meson in-medium properties in Sec.\ \ref{RD}. Finally a
summary is given in Sec.\ \ref{Sum}.

\section{The $\eta$N interactions in Chiral perturbation theory}
\label{Chpt}

\subsection{The theory framework}
The interactions between pseudoscalar
mesons (pion, kaon, and eta meson) and baryons (nucleons and
hyperons) are described by the
SU(3)$_{\mathrm{L}}\times$SU(3)$_{\mathrm{R}}$ chiral Lagrangian
which can be written as
\begin{eqnarray}\label{LL}
{\mathcal{L}_{chiral} }={\mathcal{L}_{\phi} }+{\mathcal{L}_{\phi
B} }.
\end{eqnarray}
$\mathcal{L}_{\phi}$ is the mesonic term up to second chiral
order\cite{chiral01},
\begin{eqnarray}
{\mathcal{L}_{\phi} }&=&\frac{1}{4}f^{2}\textrm{Tr}
 \partial^{\mu}\Sigma\partial_{\mu}\Sigma^{\dagger}
\nonumber\\
&&
 +\frac{1}{2}f^2 B_0
  \left\{\mbox{Tr} M_{q}(\Sigma-1)+\mathrm{h.c.}\right\}.
\end{eqnarray}
The second piece of the Lagrangian in Eq.(\ref{LL}),
$\mathcal{L}_{\phi B}$,  describes the meson-baryon interactions,
and reads at lowest order\cite{chiral01}
\begin{eqnarray} \label{L1}
{\mathcal{L}}^{(1)}_{\phi B}
 &=&
\mbox{Tr} \bar{B}(i\gamma^{\mu}\partial_{\mu}-m_{\mathrm{B}})B
 +i\mbox{Tr}\bar{B}\gamma^{\mu}[V_{\mu},B]
\nonumber\\
&&
 +D\mbox{Tr} \bar{B}\gamma^{\mu}\gamma^{5}\{A_{\mu},B\} +F
 \mbox{Tr}\bar{B}\gamma^{\mu}\gamma^{5}[A_{\mu},B],
\end{eqnarray}

The next-to-leading order chiral Lagrangian for s-wave
meson-baryon interactions reads \cite{chiral}
\begin{eqnarray} \label{L2}
 \mathcal{L}^{(2)}_{\phi B}
&=&
   a_1\mbox{Tr}\bar{B} (\xi M_q\xi+\mbox{h.c.})B
  +a_2\mbox{Tr}\bar{B}B(\xi M_q\xi+\mbox{h.c.})
\nonumber\\
&&
 +a_3\mbox{Tr}\bar{B}B\mbox{Tr}(M_q\Sigma+\mbox{h.c.})
 +d_1\mbox{Tr}\bar{B}A^2B
\nonumber\\
&&
 +d_{2}\textrm{Tr}\bar{B}(v\cdot A)^{2}B
 +d_{3}\textrm{Tr}\bar{B}BA^{2}
\nonumber\\
&&+d_{4}\textrm{Tr}\bar{B}B(v\cdot A)^{2}
+d_{5}\textrm{Tr}\bar{B}B\textrm{Tr}A^{2}
\nonumber\\
& &
 +d_{6}\textrm{Tr}\bar{B}B\textrm{Tr}(v\cdot A)^{2}
 +d_{7}\textrm{Tr}\bar{B}A_{\mu}\textrm{Tr}A^{\mu}B
\nonumber\\
&&
 +d_{8}\textrm{Tr}\bar{B}(v\cdot A)\textrm{Tr}(v\cdot A)B
 +d_{9}\textrm{Tr}\bar{B}A_{\mu}BA^{\mu}
\nonumber\\
&&
 +d_{10}\textrm{Tr}\bar{B}(v\cdot A)B(v\cdot A),
\end{eqnarray}
In the above equations, $M_{q}=\mbox{diag}\{m_{q}, m_{q}, m_{s}\}$
is the current quark mass matrix, $B_0$ relates to the order
parameter of spontaneously broken chiral symmetry, the constants
$D$ and $F$ are the axial vector couplings whose values can be
extracted from the empirical semileptonic hyperon decays, the
pseudoscalar meson decay constants are equal in the
SU(3)$_{\mathrm{V}}$ limit, and denoted by $f=f_{\pi}\simeq93$
MeV, $v_{\mu}$ is the four-velocity of the heavy baryon (with
$v^2$=1), and
$ 
\Sigma
=
 \xi^2=\exp{(i\sqrt{2}\Phi/f)},
 V^{\mu}
= (\xi\partial^{\mu}\xi^{\dag}+\xi^{\dag}\partial^{\mu}\xi)/2,
 A^{\mu}
= (\xi\partial^{\mu}\xi^{\dag}-\xi^{\dag}\partial^{\mu}\xi)/(2i).
$ 
%
The $3\times 3$ matrix $B$ is the ground state baryon octet,
$m_{\mathrm{B}}$ is the common baryon octet mass in the chiral
limit and $\Phi$ collects the pseudoscalar meson octet.

The next-to-leading-order terms in Eq.\ (\ref{L2}) have been developed for
heavy baryons by Jenkins and Manohar \cite{chiral1}. The heavy
baryon chiral theory is similar to the non-relativistic
formulation of baryon chiral perturbation theory \cite{chiral2}.
However, the heavy baryon theory has the advantage of manifest
Lorentz invariance, and quantum corrections can be computed in a
straightforward manner by the ordinary Feynman graphs, rather than
the time ordered perturbation theory \cite{chiral11}. The
Lagrangian has been shown to be suitable for describing the chiral
properties of nuclear system in Ref.\ \cite{chiral3}, where one
can also find detailed discussions on how to systematically
compute the higher order terms of this Lagrangian. In this paper,
we limit our calculations up to the squared characteristic small
momentum scale $Q^{2}$ (involving no loops) for s-wave
$\eta\mbox{N}$ scattering, because the corrections from the
higher-order coupling are suppressed, at low energy, by powers of
$Q/\Lambda_{\chi}$ with $\Lambda_{\chi}\sim 1$ GeV being the
chiral symmetry breaking scale. Hence no loops need to be
calculated in this paper. If the loop corrections are included,
the higher order terms, i.e., next-to-next-to-leading order,
should be added. We will consider it in our later work.

Expanding $\Sigma$ up to the order of $1/f^2$, and using the
heavy-baryon approximation, i.e.,
\begin{equation} \label{vapp}
v=\frac{p}{m} =\left(
  \sqrt{1+\frac{\textbf{p}^2}{m^2}},  v_x ,  v_y ,  v_z
 \right)
\approx (1,0,0,0)
\end{equation}
(because $v_x $, $v_y $ and $v_z $ are very small), we easily
obtain the Lagrangian for $\eta$N interactions:
\begin{eqnarray} \label{teffL}
\mathcal{L_{\eta }}  &=&
 \frac{1}{2}\partial^{\mu}\eta\partial_{\mu}\eta
 -\frac{1}{2}\left(
  m_{\eta}^2
-\frac{\Sigma_{\eta
\mathrm{N}}}{f^2}\bar{\Psi}_{\mathrm{N}}\Psi_{\mathrm{N}}
 \right) \eta^2\nonumber\\
 &&+\frac{1}{2}\cdot\frac{\kappa}{f^2}\bar{\Psi}_{\mathrm{N}}\Psi_{\mathrm{N}}\partial^{\mu}\eta\partial_{\mu}\eta,
\end{eqnarray}
where $m_{\eta}$ corresponds to the mass  of $\eta$-meson, which
is determined by $m_{\eta}^2=\frac{2}{3}B_0 (m_q+2m_s)$.
$\Sigma_{\eta\mathrm{N}}$ is the $\eta\mbox{N}$ sigma term, which
is determined by
\begin{eqnarray} \label{sigmaexp}
\Sigma_{\eta\mathrm{N}}
=-\frac{2}{3}[a_{1}m_{q}+4a_{2}m_{s}+2a_{3}(m_{q}+2m_{s})].
\end{eqnarray}

From Eq.(\ref{teffL}), we can see that the last three terms of
Eq.(\ref{L1}) do not contribute to the $\eta$N interactions. The
$\Sigma_{\eta \mathrm{N}}/f^2$ term in Eq.(\ref{teffL}) is deduced
from the first three terms of Eq.(\ref{L2}), which corresponds to
the chiral breaking and shifts the effective mass of $\eta$-meson
in the nuclear medium. The last term of Eq.(\ref{teffL}) is the
contribution from the last ten terms of Eq.(\ref{L2}), sometimes,
which is called ``off-shell'' term. $\kappa$ is a constant
relevant to $d_i$'s ($i=1\mbox{---}10$). Its value is to be
determined from the $\eta$N scattering length.

\subsection{The determination of the $\eta$N sigma term and $\kappa$}
To calculate $\Sigma_{\eta\mathrm{N}}$, we should know the
parameters on the right hand side of Eq.\ (\ref{sigmaexp}). In
fact these parameters have been previously discussed in Refs.\
\cite{xis,xis1,xis2,xis3} and are used in Ref.\ \cite{chiral01}.

As is well known, the KN sigma term can be written as \cite{CHL}
\begin{equation} \label{sgKN}
\Sigma_{\mathrm{KN}}
=-(m_{s}+m_{q})(a_{1}+2a_{2}+4a_{3})/2.
\end{equation}
Solving $a_3$ from this equation, and then substituting
the corresponding expression into Eq.\ (\ref{sigmaexp}) leads to
\begin{equation} \label{sigexp2}
\Sigma_{\eta\mathrm{N}} =\frac{2}{3}
 \left[
  \frac{2+r}{1+r}\Sigma_{\mathrm{KN}}
  +a_1m_s\left(1-\frac{r}{2}\right)-a_2m_s(2-r)
 \right],
\end{equation}
where $r=m_q/m_s\ll 1$. Expanding the right
hand side of Eq.\ (\ref{sigexp2}) to a Taylor series
with respect to $r$, we have
\begin{eqnarray}
\Sigma_{\eta\mathrm{N}}
&=&
 \frac{2}{3}
  \left(
   2\Sigma_{\mathrm{KN}} +a_1m_s -2a_2m_s
  \right)
\nonumber\\
&&
 -\frac{1}{3}
  \left(
   2\Sigma_{\mathrm{KN}} +a_1m_s -2a_2m_s
  \right) r
\nonumber\\
&&
 +\mbox{higher-order terms in}\ r.
\end{eqnarray}
Because of the extreme smallness of $r$, and also due to the fact
that our formulas are valid merely up to the next-to-leading
order, we take only the first two terms, i.e, $
\Sigma_{\eta\mathrm{N}} =(1/3)(2\Sigma_{\mathrm{KN}} +a_1m_s
-2a_2m_s)(2-r). $ Usually, $r$ is in the range of (1/24, 1/26)
\cite{r0,r1,r2,r3}, and we use the modest value $r=1/25$. In fact,
the concrete value does not matter significantly due to the
extreme smallness of $r$.
The values for $a_{1}m_{s}$ and $a_{2}m_{s}$ can be well
determined by Gell-Mann Okubo mass formulas, giving the result
$a_{1}m_{s}=-67$ MeV. For $a_2m_s$, one has 125 MeV \cite{chiral}
or a little bigger value 134 MeV \cite{xis3}, and we take the
average $a_{2}m_{s}=130$ MeV.
The value for KN sigma term has some uncertainties. The latest
result is $\Sigma_{\mathrm{KN}}=312\pm 37$ MeV in the perturbative
chiral quark model\cite{KN}. The lattice gauge simulation gave
$\Sigma_{\mathrm{KN}}=450\pm 30$ MeV\cite{xis5}. The result  of
lattice QCD is $\Sigma_{\mathrm{KN}}=362\pm 13$ MeV \cite{KN1},
and prediction of the Nambu-Jona-Lasinio model is
$\Sigma_{\mathrm{KN}}=425$ (with an error bar of
$10-15$\%)\cite{KN2}. Thus, in our calculations, we use
$\Sigma_{\mathrm{KN}}=380\pm 100$ MeV in its possible range.
Equipped with the above parameters, we finally obtain
$\Sigma_{\eta\mathrm{N}}=283\pm 131\ \mbox{MeV}$, where $\pm 131$
MeV reflects the uncertainty $\pm 100$ MeV in
$\Sigma_{\mathrm{KN}}$. Naturally, if one uses a smaller
$\Sigma_{\mathrm{KN}}$ vakue, e.g.,
$\Sigma_{\mathrm{KN}}=2m_{\pi}$ \cite{chiral},
$\Sigma_{\eta\mathrm{N}}$ would also become smaller.

For the other parameter $\kappa$, it is not too difficult, from
the Lagrangian in Eq.\ (\ref{teffL}), to derive the $\eta$N
scattering length (on-shell constraints):
\begin{eqnarray}
a^{\eta\mathrm{N}} =\frac{1}{4\pi f^2(1+m_{\eta}/M_{\mathrm{N}})}
 \left(\Sigma_{\eta\mathrm{N}}
+ \kappa m_{\eta}^2\right).
\end{eqnarray}
So we can determine $\kappa$ with a given
$\Sigma_{\eta\mathrm{N}}$ and $a^{\eta\mathrm{N}}$ via the relation
\begin{eqnarray} \label{akap}
\kappa =4\pi f^2
 \left(
  \frac{1}{m_{\eta}^2}+\frac{1}{m_{\eta}M_{\mathrm{N}}}
 \right)
a^{\eta\mathrm{N}} -\frac{\Sigma_{\eta\mathrm{N}}}{m_{\eta}^2}.
\end{eqnarray}

Recently, Green \emph{et al.} \cite{sc} analyzed the new
experimental data from GRAAL \cite{gall}, and gave the real part
of $\eta\mbox{N}$ scattering length  $a^{\eta\mathrm{N}}=0.91$ fm,
which agrees to their previous result \cite{sc0}. With the similar
method, Arndt \emph{et al}.\ \cite{ttt} also predicted
$a^{\eta\mathrm{ N} }=1.03-1.14$ fm, comparable to that found by
Green \emph{et al}. So one can assume that $a^{\eta\mathrm{N}}$ is
in the range of 0.91 $\sim$ 1.14 fm. Using the central value
$a^{\eta\mathrm{N}}=1.02\ \mbox{fm}$ leads to $
 \kappa=0.40 \pm 0.08
$ fm. For the eta and nucleon masses, we use $m_{\eta}=547.311$
MeV \cite{mass} and $M_{\mathrm{N}}= 939$ MeV.

It should be pointed out that the $\eta$N interactions in the
present model come from the term of $\Sigma_{\eta N}/f^2$ and the
off-shell term, while the leading Tomozawa-Weinberg term simply
vanishes. We do not consider any other non-diagonal coupled
channel, which was investigated with the chiral coupled channel
model by Waas and Weise \cite{b3}. According to their
calculations, the contribution of non-diagonal coupled channel to
the $\eta\mathrm{N}$ optical potential is on the order of $\sim
20$ MeV at normal nuclear density.

\section{In-medium properties of $\eta$-meson}
\label{mp}

The  Lagrangian for one  $\eta$-meson in nuclear matter is given
by
\begin{eqnarray}\label{lll}
{\mathcal{L}}={\mathcal{L}_{0}}+{\mathcal{L_{\eta }}},
\end{eqnarray}
where $\mathcal{L}_{0}$ is the Lagrangian for the nucleon system.
In this paper, we adopt the standard Lagrangian,
$\mathcal{L}_{0}$, for the nucleon system in relativistic
mean-field theory (given in the appendix). $\mathcal{L_{\eta }}$
is the Lagrangian for $\eta$-meson, which is given in Eq.\
(\ref{teffL}). On application of the Lagrangian in Eq.\
(\ref{lll}), we immediately have the equation of motion for the
$\eta$-meson field
\begin{eqnarray}
 \left(
  \partial_{\mu}\partial^{\mu}
  +m_{\eta}^2
  -\frac{\Sigma_{\eta N}}{f^2}\bar{\Psi}_{\mathrm{N}}\Psi_{\mathrm{N}}
  +\frac{\kappa}{f^2}\bar{\Psi}_{\mathrm{N}}\Psi_{\mathrm{N}}\partial_{\mu}\partial^{\mu}
 \right)
 \eta
 = 0.
\end{eqnarray}

Defining the $\bar{\Psi}_{\mathrm{N}}\Psi_{\mathrm{N}}$
fluctuation $\delta$ as
\begin{eqnarray}
\bar{\Psi}_{\mathrm{N}}\Psi_{\mathrm{N}}=\langle
\bar{\Psi}_{\mathrm{N}}\Psi_{\mathrm{N}}\rangle +\delta,
\end{eqnarray}
where $\langle\bar{\Psi}_{\mathrm{N}}\Psi_{\mathrm{N}}\rangle$ is
the vacuum expectation value.
Because the mean-field approximation is a very familiar method
which has already been used in studying the in-medium properties
of kaons with a similar chiral approach \cite{sc3,ch},
We adopt it in our present calculations.

At the mean-field level, we neglect the fluctuation $\delta$. Then
the equation of motion for the $\eta$-meson field is simplified to
\begin{equation} \label{simpeq}
\left(
 \partial_{\mu}\partial^{\mu}
 +m_{\eta}^2
 -\frac{\Sigma_{\eta\mathrm{N}}}{f^2}\rho_{s}
 +\frac{\kappa}{f^2}\rho_{s}\partial_{\mu}\partial^{\mu}
\right)
\eta
=0,
\end{equation}
where $\rho_{s}\equiv\langle
\bar{\Psi}_{\mathrm{N}}\Psi_{\mathrm{N}} \rangle$ is the scalar
density.

Plane wave decomposition of Eq.\ (\ref{simpeq}) yields
\begin{eqnarray} \label{decom}
-\omega^2+\vec{k}^2+m_{\eta}^2
-\frac{\Sigma_{\eta\mathrm{N}}}{f^2}\rho_{s}
+\frac{\kappa}{f^2}\rho_{s}
 \left(-\omega^2+\vec{k}^2\right)=0.
\end{eqnarray}
The $\eta$-meson effective mass, $m_{\eta}^*$, in the nuclear
medium is defined by
\begin{equation} \label{effmetadef}
\omega=\sqrt{{m_{\eta}^*}^2 +\vec{k}^2}.
\end{equation}
Substituting this equation into Eq.\ (\ref{decom})
leads to an explicit expression
\begin{eqnarray}
m_{\eta}^*
=\sqrt{
  \left(m_{\eta}^2-\frac{\Sigma_{\eta\mathrm{N}}}{f^2}\rho_s\right)
  \Big/ \left(1+\frac{\kappa}{f^2}\rho_s\right)
      }.
\end{eqnarray}

Simultaneously, the last two terms on the right hand side of Eq.\
(\ref{decom}) is the $\eta$-meson self-energy, i.e.,
\begin{eqnarray}
\Pi(\omega,\vec{k};\rho_s)
=-\frac{\Sigma_{\eta\mathrm{N}}}{f^2}\rho_{s}
 +\frac{\kappa}{f^2}\rho_{s}
  \left(-\omega^2+\vec{k}^2\right),
\end{eqnarray}
which is a function of the $\eta$-meson single-particle energy
$\omega$ and the momentum $\vec{k}$. Accordingly, the optical
potential for $\eta$-meson in the nuclear matter is given by
\begin{equation}
U_{\eta} =\frac{1}{2m_{\eta}}\Pi(\omega,\vec{k}=0;\rho_{s})
=\frac{{m_{\eta}^*}^2-m_{\eta}^2}{2m_{\eta}}.
\end{equation}

To obtain the  $\eta$-meson in-medium properties, we need a
relation between the scalar density $\rho_{s}$ and the nucleon
density $\rho_{\mathrm{N}}=\langle
\Psi_{\mathrm{N}}^{\dagger}\Psi_{\mathrm{N}}\rangle$. Because
there is only one single $\eta$-meson in the nuclear matter, its
effect on the nuclear matter is neglectable. According to the
relativistic mean-field theory, we have the following relation
between $\rho_s$, $\rho_{\mathrm{N}}$, and the $\sigma$ mean-field
value $\sigma_0$:
\begin{equation} \label{rhosx}
\rho_s
= \left(
   M_{\mathrm{N}}+g_{\sigma}^{\mathrm{N}}\sigma_0
  \right)^3 f(x),
\end{equation}
where the function $f(x)$ is defined to be
\begin{equation}
f(x)\equiv
 \left[
  x\sqrt{1+x^2}-\ln\left(1+\sqrt{1+x^2}\right)
 \right]/\pi^2
\end{equation}
with $x$ being the ratio of the nucleon's Fermi momentum to its
effective mass, i.e.,
\begin{equation} \label{xdef}
x\equiv\frac{k_{\mathrm{F}}}{M_{\mathrm{N}}^*}
=\left(\frac{3}{2}\pi^2\rho_{\mathrm{N}}\right)^{1/3}
 \Big/\left(M_{\mathrm{N}}+g_{\sigma}^{\mathrm{N}}\sigma_0\right).
\end{equation}
The mean-field value $\sigma_0$  
is connected to the scalar density $\rho_s$
by 
\begin{equation} \label{rhossigma0}
\rho_s
=
-\left(
  m_{\sigma}^{2} \sigma_{0}+g_{2}\sigma_{0}^2+g_{3}\sigma_{0}^3
 \right)
/g_{\sigma}^{\mathrm{N}}.
\end{equation}
Therefore, for a given nucleon density $\rho$, we can first
solve $\sigma_0$ from
\begin{equation} \label{eqmsigrhos}
m_{\sigma}^{2} \sigma_{0}+g_{2}\sigma_{0}^2+g_{3}\sigma_{0}^3
=-g_{\sigma}^{\mathrm{N}}
 \left(
  M_{\mathrm{N}}+g_{\sigma}^{\mathrm{N}}\sigma_0
 \right)^3 f(x),
\end{equation}
and then calculate the scalar density $\rho_s$
from Eq.\ (\ref{rhossigma0}) or (\ref{rhosx}).

\begin{figure}[ht]
\centering
\includegraphics[width=8.2cm]{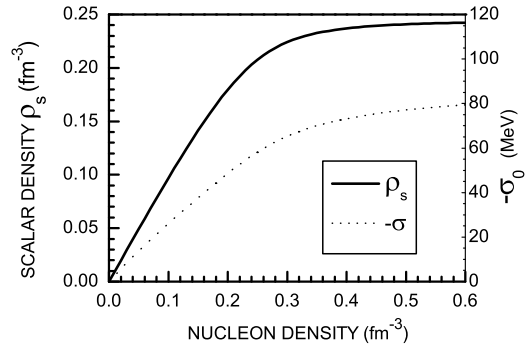}
\caption{
The scalar density (full curve) and the negative sigma mean-filed
value (dotted curve) as functions of the nucleon density. They both
are increasing functions, but the increasing speed is getting
slower and finally when the density is higher than about 2 times the
nuclear saturation density, they are nearly constant.
        }
\label{rhosrho}
\end{figure}

 The detailed derivation of the Eqs.\ (\ref{rhosx})--(\ref{eqmsigrhos})
can be seen in Ref.\ \cite{rho}. To be self-contained, we also
attach a brief derivation in the appendix. In numerical
calculations, we adopt the NL3 parameter set
\cite{Lalazissis97prc55} i.e., $m_{\sigma}=508.194$ MeV,
$m_{\omega}=782.501$ MeV, $g_{\sigma}^{\mathrm{N}}=10.217$,
$g_{\omega}^{\mathrm{N}}=12.868$, $g_2=-10.434\ \mathrm{fm}^{-1}$
and $g_3=-28.885$. The numerical results for
$\rho_s$-$\rho_{\mathrm{N}}$ are given in Fig.\ \ref{rhosrho},
where one can  see clearly that $\rho_s$ is an increasing function
of the nuclear density. When the density is  about 1.5 times lower
than the nuclear saturation density, $\rho_s$ is nearly
proportional to $\rho_{\mathrm{N}}$. However, when the density is
 about 2 times higher than the normal nuclear density,  $\rho_s$ is
nearly a constant. The mean-field value of the sigma filed is also
given in Fig.\ \ref{rhosrho} with a dotted curve. Its density
behaviour is similar to that of $\rho_s$.

\section{Results and discussions}
\label{RD}

In this section, we discuss the effective mass, optical potential
in  nuclear medium, and the off-shell behavior of $\eta$-meson,
respectively. For zero momentum $\eta$-meson, we can see, from
Eq.\ (\ref{effmetadef}), that the energy $\omega$\ is equal to its
effective mass. Therefore, we do not mention the $\eta$-meson
energies any more in the following discussions.

\begin{table}[ht]
\begin{tabular}{|c|c|c|c| }\hline\hline
Reaction or method & $a_{\eta\mathrm{N}}$ (fm)
           & $m^{*}_{\eta}$/$m_{\eta}$ & $-U_{\eta}$ (MeV) \\ \hline
\cite{s25} & 0.25 & 0.952  & 26    \\
\cite{s27} & 0.27 & 0.946 & 27    \\
$pn\longrightarrow d \eta$ \cite{s30} & $\leq$ 0.3 & $\geq$0.94 & $\leq$30  \\
\cite{sl46}& 0.46(9) & 0.915 & 44  \\
\cite{s48} & 0.487 & 0.91 & 46    \\
\cite{s51} & 0.51  & 0.905& 49 \\
\cite{s55} & 0.55  & 0.902 & 51  \\
\cite{s48} & 0.577 & 0.90 & 54   \\
\cite{s62} & 0.621 & 0.89 & 57 \\
\cite{s68} & 0.68  & 0.88 & 61  \\
\cite{s71} & 0.717 & 0.88& 63   \\
Coupled K matrices \cite{s75}& 0.75    & 0.875&67 \\
$ \eta d\longrightarrow  \eta d$ \cite{sd75}& $\geq$ 0.75   & $\leq$ 0.875&$\geq$ 67  \\
Coupled K matrices \cite{sc0}& 0.87 & 0.86&76 \\
\cite{s91,sc}& 0.91 & 0.853&77   \\
\cite{s98}& 0.98 &0.846 &79  \\
\cite{s99}& 0.991    & 0.845&80 \\
Coupled K matrices \cite{sc0}&1.05 & 0.82&82 \\
\cite{ttt}& 1.14 & 0.825&88  \\ \hline \hline
 \end{tabular}
\caption{A selection of the real part of the $\eta$N scattering
length in literature. $m^*_{\eta}$/$m_{\eta}$ and $U_\eta$ are
effective mass and the potential depth at normal nuclear density
calculated with the scattering lengths. Where we use $\Sigma_{\eta
N}=280$ MeV in calculations.} \label{paraTaba}
\end{table}

In the calculation, the precision of the $\eta$-meson effective
mass and optical potential are determined by the two parameters
$\Sigma_{\eta\mathrm{N}}$ and $\kappa$. Equation (\ref{akap})
connects the parameter $\kappa$\ to the scattering length
$a^{\eta\mathrm{N}}$, whose possible values are collected in Tab.\
\ref{paraTaba}. To reflect uncertainties in the two quantities
$\Sigma_{\eta\mathrm{N}}$ and $a^{\eta\mathrm{N}}$, we take the
sigma term $\Sigma_{\eta N}=$ 150, 280, and 410 MeV, and the
scattering length $a^{\eta\mathrm{N}}=0.91$ \cite{sc} and 1.04
\cite{ttt} fm, in numerical calculations.

Figs. \ref{S13nb} and \ref{S12nb} show  the $\eta$-meson effective
mass and nuclear optical potential of $\eta$-meson as functions of
the nuclear density. The results from Ref.\ \cite{b3} (strait
line) is also shown in Fig.\ \ref{S13nb} for comparison. The
curves in Figs.\ \ref{S13nb} and \ref{S12nb} are obviously divided
into three groups which correspond to different scattering lengths
$a^{\eta\mathrm{N}}=$ 0.91, 1.14 fm and $\kappa=0$, respectively.
The dotted, solid and dash-dotted curves in each group correspond
to $\Sigma_{\eta\mathrm{N}}=$ 150, 280, and 410 MeV, respectively.

\subsection{Effective mass}

\begin{figure}[htb]
\epsfxsize=8.2cm \epsfbox{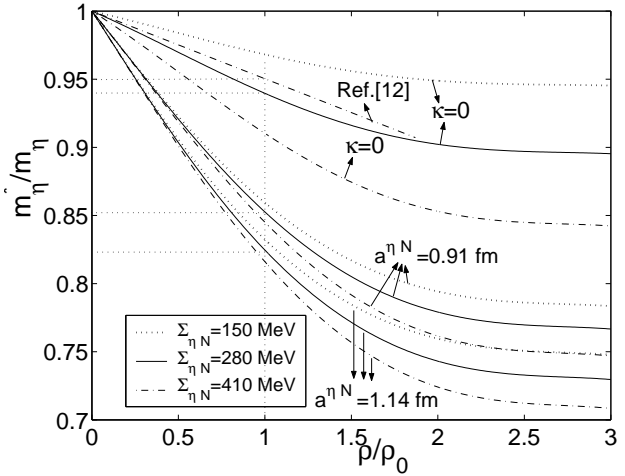} \caption{
  The Effective mass of $\eta$-meson as a function of nuclear density.
The straight line is obtained from Ref.\ \cite{b3}.
         }
\label{S13nb}
\end{figure}

It is obvious, from Fig.\ \ref{S13nb}, that the $\eta$-meson
effective mass decreases almost linearly in the region
$\rho<\rho_0$. In this region, the results of Ref.\ \cite{b3} also
show a linear relation for the effective masse with nuclear
density. At higher densities, however, the effective mass
decreases non-linearly, and the decreasing speed becomes smaller
and smaller and at last nearly constant in the range
$\rho>2\rho_{0}$. The reason is that, when the density is higher
than about 2 times the normal nuclear saturation density, $\rho_s$
nearly is a constant (see Fig.\ 1).

For the same scattering length, we find that, at low density
region $\rho\leq 0.5\rho_0$, the effective mass is nearly
independent of the sigma term $\Sigma_{\eta\mathrm{N}}$. When we
set $\Sigma_{\eta\mathrm{N}}=280\pm 130$ MeV, which changes in a
large range, the variation of the effective mass is within $\pm 4$
MeV at $\rho=\rho_0$. And at high nuclear density, say
$\rho=3\rho_0$, the variation is within $\pm 10$ MeV compared with
that at the central value of $\Sigma_{\eta N}=280$ MeV. Thus, we
can conclude that the effective mass of $\eta$-mesons is
insensitive to the concrete value of $\Sigma_{\eta\mathrm{N}}$ at
low density region.

Although the latest predictions \cite{sc,ttt} give large
scattering lengths $a^{\eta N}=0.91\sim 1.14$ fm, there are other
different predictions
\cite{s25,s27,s30,sl46,s48,s51,s55,s62,s68,s71,s75,sd75,sc0,s91,s98,s99}.
To see the effects of different scattering length values on the
$\eta$\ effective mass, we show, in Fig.\ \ref{S13nb}, the results
for $a^{\eta N}=0.91$ and 1.14 fm, respectively. On the other
hand, in Tab. I, we give, at normal nuclear density, the effective
mass corresponding to the respective $\eta$N scattering length in
the literature
\cite{s25,s27,s30,sl46,s48,s51,s55,s62,s68,s71,s75,sd75,s91,s98,s99,sc0,sc,ttt}.
From Fig.\ \ref{S13nb}, we find that, with the same sigma term
$\Sigma_{\eta N}$, the effective mass depends strongly on the
scattering length $a^{\eta N}$. At $\rho=\rho_{0}$, the effective
mass (with $\Sigma_{\eta N}=280$MeV) is $m^*_{\eta}/m_{\eta}=0.85$
for $a^{\eta N}=0.91$ and $m^*_{\eta}/m_{\eta}=0.825$ for $a^{\eta
N}=1.14$ fm. When the scattering length varies from 0.25 fm to
1.14 fm, the effective mass will run from 0.95$m_{\eta}$ to
0.825$m_{\eta}$. Corresponding to $a^{\eta N}=0.91\sim 1.14$ fm,
which are favored by recent works, and $\Sigma_{\eta\mathrm{N}}$,
which is predicted in Sec.\ II, the effective mass is $(0.84\pm
0.015)m_{\eta}$.

At normal nuclear density, the effective mass in Ref.\ \cite{b3}
is 0.95$m_{\eta}$, which agrees with result of the small
scattering length $a^{\eta N}=0.25$ fm. As pointed out in the
above, the effective mass changes nonlinearly with increasing
densities in the region $\rho_0<\rho<2\rho_0$. This behavior
agrees to the predictions by Tsushima \emph{et al.}\ \cite{efm}
with quark-meson coupling model. The effective mass at
$\rho=\rho_0$ predicted by them is about 0.88$m_{\eta}$, which
just corresponds to the result with scattering length $a^{\eta
N}=0.68$ fm. This can be clearly seen from Tab.\ I . The
outstanding characteristic of our results is that the present
calculations give much smaller effective mass than the others when
we adopt the larger scattering length.

It should be mentioned that the chiral coupled channel model
\cite{b3} gives much larger in-medium effective mass for
$\eta$-mesons than our predictions. The main reason is as such. In
the chiral coupled channel model, there are only the leading-order
terms, and so, the contributions to the effective mass  come only
from the non-diagonal coupled channel. While in our model, the
leading-order terms do not contribute to the calculations. All the
contributions to the results come from the next-to-leading-order
terms.

\subsection{Optical potential}

\begin{figure}[htb]
\epsfxsize=8.2cm \epsfbox{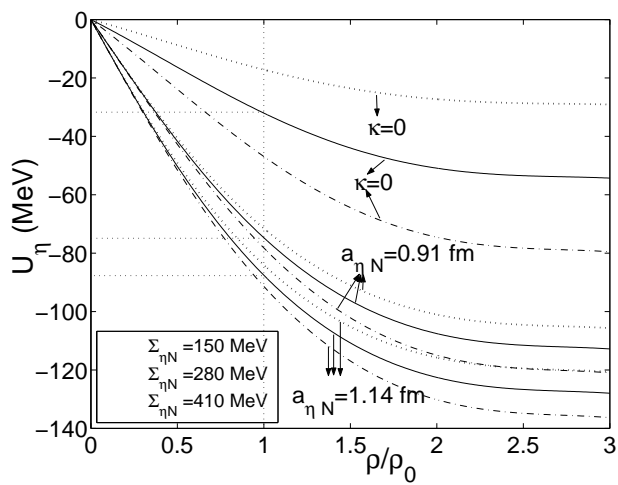} \caption{
 The optical potential of $\eta$-mesons as a function of the nuclear
density.
         }
\label{S12nb}
\end{figure}

The optical potential $U_{\eta}$ as a function of nuclear density
is plotted in Fig.\ \ref{S12nb}. We find that the density behavior
of $U_{\eta}$ is quite similar to the effective mass in Fig.\
\ref{S12nb}. The reason is that the optical potential has a
relation $U_{\eta}\simeq m_{\eta}^*-m_{\eta}$ as an approximation,
which varies linearly with the effective mass $ m_{\eta}^*$ of
$\eta$-meson.

Similarly, it is also seen that the effect from the uncertainties
of sigma term $\Sigma_{\eta N}$ are quite limited in its possible
range, and the optical potential depend strongly on the value of
the scattering length. At normal nuclear density, the upper limit
of the uncertainties from the sigma term $\Sigma_{\eta N}$ is no
more than $8$ MeV. However, the optical potential can change from
$-78$ MeV to $-88$MeV, when we modify the scattering length
$a^{\eta N}$ from 0.91 to 1.14 fm. Because there are still
uncertainties for the $\eta$N scattering length, we listed the
possible potential depths corresponding to the possible scattering
lengths appearing in literature in Tab.\ I. From the table, we can
see that the potential depth at normal nuclear density ranges from
26 MeV to 88 MeV, because of the uncertainties of scattering
lengths. According to the newest predictions, i.e.\ $a^{\eta N}=
0.91 \sim 1.14$ fm\cite{sc,ttt}, the potential depth is about
$83\pm 5$ MeV. This is a very strong attractive potential which
was never predicted by the previous models.

There have been some predictions for the nuclear potential of
$\eta$-mesons in other references. According to the SU(3) chiral
dynamics with coupled channels, the optical potential depth at
normal nuclear density is $U_{\eta}\simeq-20$ MeV \cite{b3}, which
is close to our formulas with a smaller scattering length $a_{\eta
N}< 0.25$ fm. In Ref.\ \cite{op}, by assuming that the mass of the
$N^*(1535)$ did not change in the medium, the optical potential
$U_{\eta}=-34$ MeV was obtained, which is close to our calculation
with $a_{\eta N}\sim 0.30$ fm. The $\eta$\ potential from QMC
model by Tsushima \emph{et al.} and chiral unitary approach by
Inoue \emph{et al.}, are typically $-60$ MeV and $-54$ MeV, which
are comparable to our formulas with $a_{\eta N}= 0.55-0.68$ fm.
Therefore, if we want to obtain shallower optical potential, we
need to use a smaller scattering length. Because recent works
favor the bigger scattering length, our formulas give much deeper
optical potential.

\subsection{The effect of off-shell term }

Finally, we discuss the role of the off-shell term in our
calculation. In present model, the off-shell term $\kappa$ is
determined by the scattering length $a^{\eta N}$. From the
analysis of the subsections A and B, we know that the scattering
length $a^{\eta\mathrm{N}}$ strongly affects the calculations. The
importance of the off-shell behavior for low energy scattering had
been pointed out in many Refs.\cite{off,chiral,sc3}.

To clarify the effect of off-shell term on our calculation
thoroughly, we turn off the off-shell term ($\kappa=0$) and show
the results in Fig.\ \ref{S13nb} and \ref{S12nb}. At
$\rho=\rho_{0}$, without the off-shell terms, the effective mass
is $m^*_{\eta}/m_{\eta}\simeq 0.94\pm0.03$, and the optical
potential is $-(32\pm 16)$ MeV corresponding to $\Sigma_{\eta
N}=280\pm 130$ MeV. Thus, without the off-shell terms, we no
longer have strong attractive potential for $\eta$-meson in
nuclear medium. The calculations are independent of the scattering
length. Also in this case, the calculations depend strongly on the
quantity of $\Sigma_{\eta N}$. Without the off-shell terms, the
variation of the optical potential from the uncertainties of
$\Sigma_{\eta N}$ can reach about 30 MeV at normal nuclear
density. However, it is no more than 8 MeV, when the off-shell
behavior is considered. Thus, the off-shell terms can dramatically
depress the effects from the uncertainties of $\Sigma_{\eta N}$.

\section{Summary}
\label{Sum}

In this paper, we have derived an effective Lagrangian for $\eta$N
s-wave interaction from the effective meson-baryon chiral
Lagrangian including the next-to-leading-order terms. Up to
$1/f^2$ terms for s-wave $\eta$N interaction, only the sigma term
and off-shell term survive. It is found that the $\eta$N sigma
term is $\Sigma_{\eta N}=280\pm130$ MeV according to the KN sigma
term. The off-shell term $\kappa$ is determined by the scattering
length. If we adopt the newest predictions $a_{\eta N}\approx
0.91-1.14$ fm for the scattering lengths\cite{sc,ttt}, we obtain
the value $\kappa=0.40\pm0.08$ fm.

Combining the relativistic mean-field theory for nucleon system,
we calculate the effective mass and optical potential of
$\eta$-mesons in uniform nuclear medium in the mean-field
approximation. According to the latest predictions $a_{\eta
N}\approx 0.91-1.14$ fm for the scattering lengths\cite{sc,ttt},
at normal nuclear density the effective mass is about $(0.84\pm
0.015)m_{\eta}$, and the depth of optical potential is
$U_{\eta}\simeq -(83\pm 5)$ MeV.

Finally, we should point out the importance of the
next-to-leading-order terms of the chiral Lagrangian again. In
fact, the leading-order terms do not contribute to the $\eta$N
interactions. All contribution comes from the next-to-leading
order terms. It indicates that the next-to-leading order terms
should be included in the study of the $\eta$N interaction. In the
present paper, we do not consider corrections from the
non-diagonal coupled channel. According the study of Waas and
Weise, the correction may be on the order of 20 MeV for the
optical potential.

\section*{  Acknowledgements }

This work was supported  by the Natural Science Foundation of
China (10275037, 10375074, 10575054, 90203004) and the Doctoral
Programme Foundation of the China Institution of Higher Education
(20010055012). We thank Prof.\ X.\ B.\ Zhang for helpful
discussions.

\appendix

\section{relation between the scalar density and nucleon density
in the relativistic mean-field approach}

In this appendix, we give a short derivation of the relation
between the scalar density and nucleon density
in the relativistic mean-field approach (RMF).

In RMF, the effective Lagrangian density \cite{rmf} can be written
as
\begin{eqnarray}
\mathcal{L}_{0} &=& \bar{\Psi}_{\mathrm{N}} (
i\gamma^{\mu}\partial_{\mu}
 -M_\mathrm{N})\Psi_\mathrm{N}-g_{\sigma}^N\bar{\Psi }_\mathrm{N}\sigma\Psi_\mathrm{N}
\nonumber\\
& &
 -g_{\omega}^N\bar{\Psi}_\mathrm{N}\gamma^{\mu}\omega_{\mu}\Psi_\mathrm{N}
 -g_{\rho}^{N} \bar{\Psi}_{\mathrm{N}}
  \gamma^{\mu}\rho^a_{\mu}\frac{\tau_a}{2}\Psi_\mathrm{N}
\nonumber\\
& &
 +\frac{1}{2}\partial^{\mu}\sigma\partial_{\mu}\sigma
 -\frac{1}{2}m_{\sigma}^{2}\sigma^{2}
 -\frac{1}{3}g_{2}^{2}\sigma^{3}
 -\frac{1}{4} g _{3}^{2}\sigma^{4}
\nonumber\\
& &
 -\frac{1}{4}\Omega^{\mu\nu}\Omega_{\mu\nu}
 +\frac{1}{2}m_{\omega}^{2}\omega^{\mu}\omega_{\mu}
 -\frac{1}{4}  R ^{a \mu\nu}R _{\mu\nu}^{a}
\nonumber\\
& &
 +\frac{1}{2} m_{\rho}^{2} {\rho^{a\mu}} {\rho^{a}_{\mu}}
 -\frac{1}{4} F^{\mu\nu}F_{\mu\nu}
\nonumber\\
& &
 -e\bar{\Psi }_{\mathrm{N}}\gamma^{\mu}A^{\mu}\frac{1}{2}(1+\tau_{3})\Psi_{\mathrm{N}}
\end{eqnarray}
with
$
\Omega^{\mu\nu}
=\partial^{\mu}\omega^{\nu}
 -\partial^{\nu}\omega^{\mu},
 \
 R^{a\mu\nu}
=\partial^{\mu}\rho^{a\nu}
 -\partial^{\nu}\rho^{a\mu},
 \
F^{\mu\nu} =\partial^{\mu}A^{\nu}-\partial^{\nu}A^{\mu}. $ On
application of the mean-field approximation, we have the equation
of motion for nucleons:
\begin{equation}
\left(
 \gamma_{\mu}k^{\mu}
 -M_{\mathrm{N}}
-g_{\sigma}^{\mathrm{N}}\sigma_{0}
 -g_{\omega}^{\mathrm{N}}\gamma^0\omega_0
 -g_{\rho}^{\mathrm{N}}\gamma^0\tau^3{\rho_0}_3
\right) \Psi_{\mathrm{N}}
=0.
\end{equation}
where the $\sigma$, $\omega$, and $\rho$ fields are
replaced with their mean-field values $\sigma_0$, $\omega_0$
and $\rho_0$.
$\sigma_0$ and $\omega_0$ satisfy
\begin{eqnarray}
m_{\sigma}^2\sigma_0+g_2\sigma_0^2+g_3\sigma_0^3
&=&
 -g_{\sigma}^{\mathrm{N}}\rho_s,  \label{eqsig02} \\
 m_{\omega}^2 \omega_0 &=& g_{\omega}^{\mathrm{N}}\rho_{\mathrm{N}},
                                         \label{eqome0}
\end{eqnarray}
with $ \rho_{s}
\equiv\langle\bar{\Psi}_{\mathrm{N}}\Psi_{\mathrm{N}}\rangle $ and
$ \rho_{\mathrm{N}} \equiv\langle
\Psi_{\mathrm{N}}^{\dag}\Psi_{\mathrm{N}}\rangle. $ Therefore, at
the mean-field level, the energy density of nuclear matter is
\begin{eqnarray} \label{Enm}
\varepsilon
&=& \frac{1}{2}m_{\sigma}^2\sigma_0^2
  +\frac{1}{3}g_2\sigma_0^3
  +\frac{1}{4} g_3\sigma_0^4
  +\frac{1}{2}m_{\omega}^2\omega_0^2
\nonumber\\
&& +\frac{4}{(2\pi)^3}\int_0^{k_{\mathrm{F}}}
   \left(\vec{k}^2+{M_{\mathrm{N}}^*}^2\right)^{1/2}\mathrm{d}\vec{k},
\end{eqnarray}
where $ M_{\mathrm{N}}^*
=M_{\mathrm{N}}+g_{\sigma}^{\mathrm{N}}\sigma_0 $ is the effective
mass of nucleons.

In Eq.\ (\ref{Enm}), the energy density has been expressed as
an explicit function of $\sigma_0$.
Because $\sigma_0$ should minimize $\varepsilon$, i.e.,
$\partial\varepsilon(\sigma_0)/\partial\sigma_0$, we immediately have
\begin{equation} \label{eqappfin}
m_{\sigma}^2\sigma_0+g_2\sigma_0^2+g_3\sigma_0^3
=-\frac{4g_{\sigma}^{\mathrm{N}}}{(2\pi)^3}\int_0^{k_{\mathrm{F}}}\!
  \frac{M_{\mathrm{N}}^*\,\mathrm{d}\vec{k}}
      {(\vec{k}^2+{M_{\mathrm{N}}^*}^2)^{1/2}},
\label{eqsimems}
\end{equation}
which is nothing but the Eq.\ (\ref{eqmsigrhos}).
Equation (\ref{eqsig02}) corresponds to Eq.\ (\ref{rhossigma0}).
And comparing Eq.\ (\ref{eqappfin}) with Eq.\ (\ref{eqsig02})
then gives the Eq.\ (\ref{rhosx}).

\end{document}